\documentclass[%
 reprint,
 amsmath,amssymb,
 aps,
]{revtex4-2}
\usepackage{array}
\newcolumntype{M}[1]{>{\centering\arraybackslash}m{#1}}
\usepackage{multirow}
\usepackage{graphicx}
\usepackage{epstopdf}
\usepackage{dcolumn}
\usepackage{bm}
\usepackage{color}
\usepackage[dvipsnames]{xcolor}
\usepackage{soul}
\newcommand{\eqN}[1]{Eq.~(\ref{#1})} 

\begin{document}

\preprint{APS/123-QED}

\title{Linear Nonreciprocal Dynamics of Coupled Modulated Systems}

\author{Jiuda Wu}
 \email{w\_jiuda@encs.concordia.ca}
\author{Behrooz Yousefzadeh}%
 \email{behrooz.yousefzadeh@concordia.ca}
\affiliation{%
Department of Mechanical Industrial and Aerospace Engineering, Concordia University\\
1455 De Maisonneuve Blvd. W. EV-4.139, Montreal, QC, H3G 1M8, CANADA
}

\date{\today}

\begin{abstract}

Waveguides subject to spatiotemporal modulations are known to exhibit nonreciprocal vibration transmission, whereby interchanging the locations of the source and receiver change the end-to-end transmission characteristics. The scenario of typical interest is unidirectional transmission in long, weakly modulated systems: when transmission is possible in one direction only. Here, with a view toward expanding their potential application as devices, we explore the vibration characteristics of spatiotemporally modulated systems that are short and strongly modulated. Focusing on two coupled systems, we develop a methodology to investigate the nonreciprocal vibration characteristics of both weakly and strongly modulated systems. In particular, we highlight the contribution of phase to nonreciprocity, a feature that is often overlooked. We show that the difference between the transmitted phases is the main contributor to breaking reciprocity in short systems. We clarify the roles of primary and side-band resonances, and their overlaps, in breaking reciprocity. We discuss the influence of modulation amplitude and wavenumber on the resonances of the modulated system. 
\end{abstract}

\maketitle

\section{\label{sec:level1}INTRODUCTION}

The principle of reciprocity states that propagation of elastic or acoustic waves in a medium remains invariant upon interchanging the positions of the source and receiver~\cite{Recip_book}. The reciprocity invariance generally holds in time-invariant materials functioning in the linear (small-amplitude) operating regime. This property has led to the development of various wave processing techniques and industrial applications, such as calibration of hydrophones and crack identification~\cite{Recip_app1,Recip_app2}.

In situations where reciprocity holds, the wave propagation properties (speed, amplitude, phase, etc.) cannot be controlled or tuned by changing the direction of propagation. Therefore, one way to enable direction-dependent vibration transmission is to circumvent reciprocity. Understanding the underlying mechanism for nonreciprocal propagation can enable the design and development of novel devices for energy harvesting, vibration isolation and signal processing. The theories and applications of nonreciprocal wave propagation have drawn attention of many researcher in recent years~\cite{NRM2020}.

Nonlinearity can break the reciprocity invariance in systems with broken mirror symmetry~\cite{nonlinear_amp1,nonlinear_amp2,nonlinear_amp3,nonlinear_amp4}.
In linear systems, changing one or more of the effective properties of the system as a function of time {\it and} space is an effective approach to break the time-reversal symmetry and enable nonreciprocal transmission~\cite{time_var}. The time- and space-varying term within an effective property of the system is called spatiotemporal modulation.

Continuous media with wavelike spatiotemporal modulations in elasticity were used to study nonreciprocal wave propagation~\cite{NJP2016,JSV2019,Kargozarfard2023,Deymier-2012,AIP2018}. Here, nonreciprocity manifests as directional bandgaps in the dispersion curves, which indicates unidirectional transmission of energy through the system~\cite{tilt}.
Inerters mounted on a vibrating base have been demonstrated to enable nonreciprocal transmission in a fully mechanical waveguide~\cite{CELLI2024}. 
Nonreciprocal transmission of bending and longitudinal vibrations have been analyzed for a beam with local modulated attachments, indicating the importance of finite size effects~\cite{bar_boundary}. 
Nonreciprocal wave propagation also occurs in a medium with two-phase modulation; {\it i.e.} when both elastic modulus and density change spatiotemporally~\cite{JMPS2017}. 
Moving media exhibit asymmetric dispersion characteristics too, including directional bandgaps~\cite{AIP2018,Hasan2023}. 

Nonreciprocity has been explored in discrete models of modulated materials as well. Unidirectional wave propagation can happen in metamaterials in which modulations are introduced to the stiffness of resonant springs~\cite{PRSA2017}, grounding springs~\cite{KARLICIC2023}, or springs of surface oscillators~\cite{PU2024118199,JMPS2020,Rayleigh2021}. A study on a modulated system with only two degrees of freedom highlighted the role of phase as a contributor to nonreciprocity~\cite{CSME2022}. Experimental studies on nonreciprocal vibration transmission due to spatiotemporal modulations were performed on setups that are discrete and finite in length, for example by using piezoelectric materials to change stiffness~\cite{piezoelectric,piezoelectric2} and tuning electromagnet forces on magnetic masses~\cite{YWang2018, YChen2019, APL2022_Wan,Kang2023}. 

In this work, our objective is to systematically investigate the influence of different system parameters on steady-state vibration transmission in very short modulated systems under weak or strong modulation amplitudes. We focus exclusively on a system with two degrees of freedom (2-DoF). This is the smallest possible system for investigating nonreciprocity. We aim to understand the role of different parameters, particularly the modulation amplitude and wavenumber, on nonreciprocal vibration transmission characteristics. The evaluation of (non)reciprocity in vibration transmission involves comparing both transmitted amplitudes (energies) and transmitted phases. Notably, the impact of strong modulations on the steady-state response of discrete systems is explored here for the first time. Section II provides an analysis of a 2-DoF modulated system, along with an introduction to the solution methodology utilized throughout the paper. In Section III, we delve into the characteristics of weakly modulated systems with a specific focus on nonreciprocity. 
Section IV presents the study of vibration transmission in strongly modulated systems. 
Phase nonreciprocity, identified in both weakly and strongly modulated systems, is elaborated upon in Section V. Section VI summarizes our findings.

\section{Analysis of a 2-DoF modulated system}

\subsection{Formulation of the problem}
\label{sec:EOM}
\begin{figure}[b]
    \includegraphics[scale=0.5]{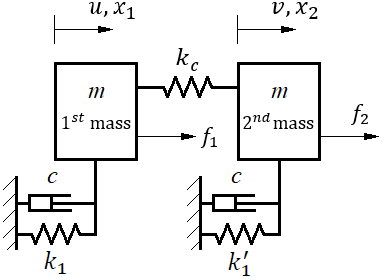}
    \caption{\label{fig:2dof} Schematic of the 2-DoF model of coupled modulated systems.}
\end{figure}
Fig.~\ref{fig:2dof} shows the schematic of the 2-DoF model of the coupled modulated systems studied in this work. 
The model consists of two identical masses, $m$, that are connected by a linear spring of stiffness $k_c$.
$u(t)$ and $v(t)$ denote the rectilinear displacement of each mass from its static equilibrium position. Each mass is grounded by a spring with a temporally modulated stiffness, as well as a linear viscous damper.
The modulated stiffness coefficients are $k_1(t)\!=\!k_{g,DC}\!+\!k_{g,AC}\cos(\omega_mt)$ and $k_1^\prime(t)\!=\!k_{g,DC}\!+\!k_{g,AC}\cos(\omega_mt\!-\!\phi)$, each with a constant component $k_{g,DC}$ and a time-dependent modulation of amplitude $k_{g,AC}$ and frequency $\omega_m$.
The parameter $\phi$ represents the phase difference between the modulations of the two grounding springs. This is equivalent to the modulation wavenumber in a spatiotemporally modulated lattice. 
External harmonic forces are applied on each mass, expressed by $f_1(t)\!=\!F_1\cos(\omega_ft)$ and $f_2(t)\!=\!F_2\cos(\omega_ft)$.

We start by nondimensionalizing the governing equations, as detailed in Appendix \ref{appendix:nondimensionalization}.
In terms of nondimensional parameters, the equations of motion for the 2-DoF modulated system are:
\begin{subequations}
\label{eq1}
\begin{eqnarray}
    \ddot{x}_1+2\zeta\dot{x}_1+[1+K_m\cos(\Omega_m\tau)]x_1\hspace{2.25cm}\nonumber\\
    \hspace{2.6cm}+K_c(x_1-x_2)=P_1\cos(\Omega_f\tau),\label{eq:1a}\\
    \ddot{x}_2+2\zeta\dot{x}_2+[1+K_m\cos(\Omega_m\tau-\phi)]x_2\hspace{1.6cm}\nonumber\\
    +K_c(x_2-x_1)=P_2\cos(\Omega_f\tau).\hspace{.06cm}\label{eq:1b}
\end{eqnarray}
\end{subequations}
Note that Eqs.~(\ref{eq:1a}) and (\ref{eq:1b}) are identical when $\phi=0$.

To investigate reciprocity, we need two configurations:
(i) the \textit{forward} configuration (or left-to-right, L2R) with $P_1=P$ and $P_2=0$, where the output is the steady-state response of the second mass, $x_2^F(\tau)$;
(ii) the \textit{backward} configuration (or right-to-left, R2L) with $P_1=0$ and $P_2=P$, where the output is the steady-state response of the first mass, $x_1^B(\tau)$.
A reciprocal response is then characterized by $x_2^F(\tau)/P_1=x_1^B(\tau)/P_2$ in this case, or simply $x_2^F(\tau)=x_1^B(\tau)$ because we use the same forcing amplitude for the \textit{forward} and \textit{backward} configurations.

The response of the modulated system is characterized by two frequencies $\Omega_f$ and $\Omega_m$.
Because these frequencies are independent from each other (incommensurate), the steady-state response of the system is neither harmonic nor periodic; it is quasi-periodic.
To characterize the quasi-periodic response in the \textit{forward} and \textit{backward} configurations, we use the output norms $N^F$ and $N^B$, respectively, which are defined as:
\begin{subequations}
\label{eq:norms0}
\begin{eqnarray}
    N^{F,B}&=\lim_{T \to \infty}\sqrt{\frac{1}{T}\int_{0}^{T} [x_{2,1}^{F,B}(\tau)]^{2} d\tau},\hspace{.7cm}\label{eq:2a} \\
    R&=\!\lim_{T \to \infty}\!\sqrt{\frac{1}{T}\!\int_{0}^{T}\! [x_{2}^{F}(\tau)\!-\!x_{1}^{B}(\tau)]^{2} d\tau}.\label{eq:2c}
\end{eqnarray}
\end{subequations}
$R$ is called the reciprocity bias, which quantifies the degree of (non)reciprocity of the system.
By definition, $R=0$ if and only if vibration transmission through the system is reciprocal.

\subsection{Solution methodology}
\label{sec:method}
In the absence of a tractable exact analytical solution to \eqN{eq1}, we use approximate methods to obtain analytical expressions for the steady-state response of the system.
Informed by the numerical observations made in Appendix~\ref{appendix:fft}, we write the steady-state response of the system in the \textit{forward} and \textit{backward} configurations as follows:
\begin{equation}
\label{eq3}
\begin{array}{l}
    \displaystyle x_{j}^{F,B}(\tau)=\sum^{\infty}_{q=-\infty} [y_{j,q}^{F,B}e^{i(\Omega_f+q\Omega_m)\tau}+c.c.] \\ 
    \displaystyle \hspace{1cm}=\sum^{\infty}_{q=-\infty} 2|y_{j,q}^{F,B}| \cos[(\Omega_f+q\Omega_m)\tau+\Psi_{j,q}^{F,B}]
\end{array}
\end{equation}
where $j\in\{1,2\}$ indicates the $1^{st}$ mass or the $2^{nd}$ mass, $c.c.$ denotes the complex conjugate terms, and $i$ is the imaginary unit.
$y_{j,q}^{F}$ and $y_{j,q}^{B}$ are the complex-valued amplitudes of the harmonic components in the steady-state response in the two configurations. 
The phase angles are $\Psi_{j,q}^F=$atan2$\,($imag$\,(y_{j,q}^F),\,$real$\,(y_{j,q}^F))$ and $\Psi_{j,q}^B=$atan2$\,($imag$\,(y_{j,q}^B),\,$real$\,(y_{j,q}^B))$.
The modal expansion in \eqN{eq3}, in addition to satisfying the numerical observations in Appendix~\ref{appendix:fft}, is a suitable asymptotic solution for either weak or strong modulations, and has been used extensively in the literature on modulated materials \cite{ASA-Haberman-2019,PRE-Haberman-2019,Ruzzene-2019,NJP2023_Ruzzene,PRA2022_Nouh,Deymier-2012,CELLI2024}.

To calculate the complex-valued amplitudes $y_{2,q}^{F}$ and $y_{1,q}^{B}$, we use the method of averaging~\cite{NayfehMook}.
The full details of this procedure are provided in Appendix~\ref{appendix:avg}.
The outcome is a linear system of algebraic equations for the unknown amplitudes in \eqN{eq3}; see \eqN{eqC3} in Appendix~\ref{appendix:avg}.
Subsequently, the expressions for the output norms and reciprocity bias in \eqN{eq:norms0} can be rewritten in terms of the complex-valued amplitudes as:
\begin{subequations}
\label{eq:norms}
\begin{eqnarray}
    N^{F,B}=\sqrt{2 \sum^{\infty}_{q=-\infty} |y_{2,1,q}^{F,B}|^{2}},\hspace{.7cm}\label{eq:4a} \\
    R=\sqrt{2\sum^{\infty}_{q=-\infty}|y_{2,q}^{F}-y_{1,q}^{B}|^{2}}.\label{eq:4c}
\end{eqnarray}
\end{subequations}

In practice, the infinite summation in \eqN{eq:norms} needs to be truncated at a finite value of $q$, for example $q\in[-\mathcal{F},\mathcal{F}]$ with $\mathcal{F}\in\mathbb{N}$, to approximate the response of the system and the output norms in \eqN{eq:norms}.
In general, the magnitude $|y_{j,q}|$ of a harmonic component (its participation in the steady-state response $x_{j}$) becomes smaller as $q$ increases, and the accuracy of the overall approximation improves by increasing the value of $\mathcal{F}$.
We choose higher values of $\mathcal{F}$ for systems with strong modulations than systems with weak modulations.
Appendix~\ref{appendix:avg} provides examples of the comparison between the results obtained from this approximation and direct numerical integration of the equations of motion.

The approximate solution developed here provides the convenience that the non-periodic steady-state response of a modulated system can be obtained by solving a linear algebraic system.
Moreover, nonreciprocity can be investigated and understood by focusing on the differences between each pair of harmonic components in the two outputs:
$2|y_{2,q}^F| \cos[(\Omega_f+q\Omega_m)\tau-\Psi_{2,q}^F]$ and $2|y_{1,q}^B| \cos[(\Omega_f+q\Omega_m)\tau-\Psi_{1,q}^B]$. We refer to such pair as a component-pair for ease of reference.
For a given $q$, the frequencies of a component-pair are the same. Either $|y_{2,q}^F|\neq|y_{1,q}^B|$ or $\Psi_{2,q}^F\neq\Psi_{1,q}^B$ can indicate a nonreciprocal response.

\section{Vibration transmission in weakly modulated systems}
\label{sec:weak}
In this section, we provide a parametric study of the steady-state nonreciprocal dynamics for weak modulations, characterized by $K_m\le0.1$. We start with investigating the effects of $K_c$ and $\Omega_m$, followed by the role of the symmetry-breaking parameter $\phi$ in breaking reciprocity.

\subsection{Primary bands and sidebands}
\label{sec:sidebands}

\begin{figure}[b]
\includegraphics[scale=0.45]{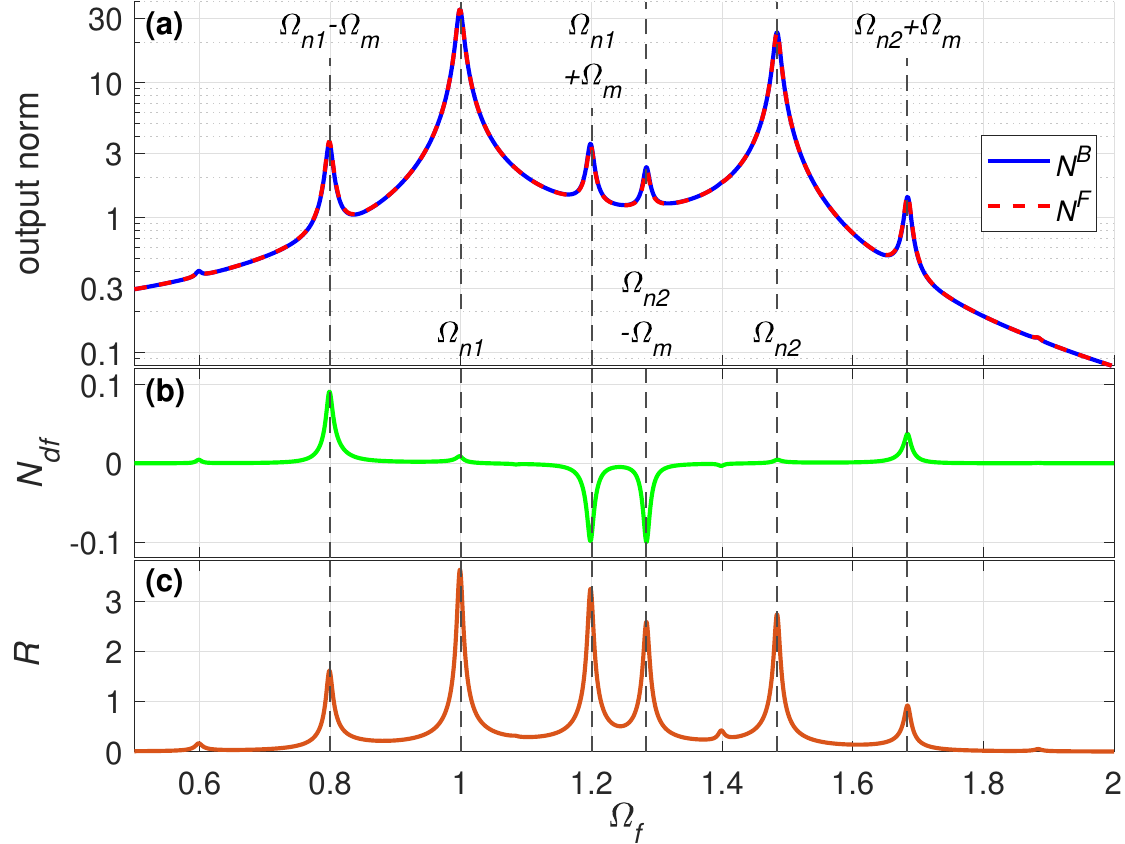}
\caption{\label{fig:NF-Omg-1} Plots of (a) output norms, (b) difference between output norms and (c) reciprocity bias as functions of $\Omega_f$. System parameters: $K_c=0.6$, $\zeta=0.005$, $K_m=0.1$, $\Omega_m=0.2$, $\phi=\pi/2$ and $P=1$.
}
\end{figure}

Fig.~\ref{fig:NF-Omg-1}(a) shows the response of the system in the \textit{forward} and \textit{backward} configurations as a function of the forcing frequency, $\Omega_f$. The natural frequencies of the unmodulated system ($K_m=0$) are $\Omega_{n1}=1$ and $\Omega_{n2}=\sqrt{1+2K_c}$. We observe primary resonances occurring at $\Omega_{n1,2}$, accompanied by sideband (secondary) resonances at $\Omega_{n1}\pm\Omega_m$ and $\Omega_{n2}\pm\Omega_m$. The values of $K_c$ and $\Omega_m$ are chosen such that there is no overlap between sideband resonances; we consider overlap in Section~\ref{sec:overlap}. 

Although we do not expect the response to be reciprocal ($\phi\ne0$), the output norms in Fig.~\ref{fig:NF-Omg-1}(a) seem to indicate a reciprocal response. Fig.~\ref{fig:NF-Omg-1}(b) shows the difference between the two output norms (transmitted energies),
\begin{equation}
    \label{eq:Ndf}
    N_{df}=N^F-N^B.
\end{equation}
We see that the transmitted energies are almost equal in the two directions, with a small difference that occurs predominantly at the sideband resonances. 

Fig.~\ref{fig:NF-Omg-1}(c) shows the reciprocity bias, $R$. It shows that despite equal energies transmitted in the opposite directions (small $N_{df}$), the transmission of vibrations is nonreciprocal ($R>0$) throughout the entire range of forcing frequencies considered. Notably, the value of $R$ is much larger than the value of $N_{df}$. Furthermore, the reciprocity bias is large not only at sideband resonances, but also at the primary resonances. 

This difference between the values of $R$ and $N_{df}$ indicates that the phase difference between output displacements is the main contributor to breaking reciprocity in this system. In other words, the phase difference between the two outputs makes a much greater contribution to nonreciprocity than the amplitude difference between them.

To better understand the contribution of phase to breaking reciprocity, we consider the amplitudes and phases of the main three harmonic components of the response; $q\in\{-1,0,1\}$ in \eqN{eq3}. Recall that because $K_m$ is small (weak modulation), the magnitudes of the outer sidebands ($|q|>1$) are increasingly small and their contribution to the overall response is negligible. 

\begin{figure}[b]
\includegraphics[scale=0.45]{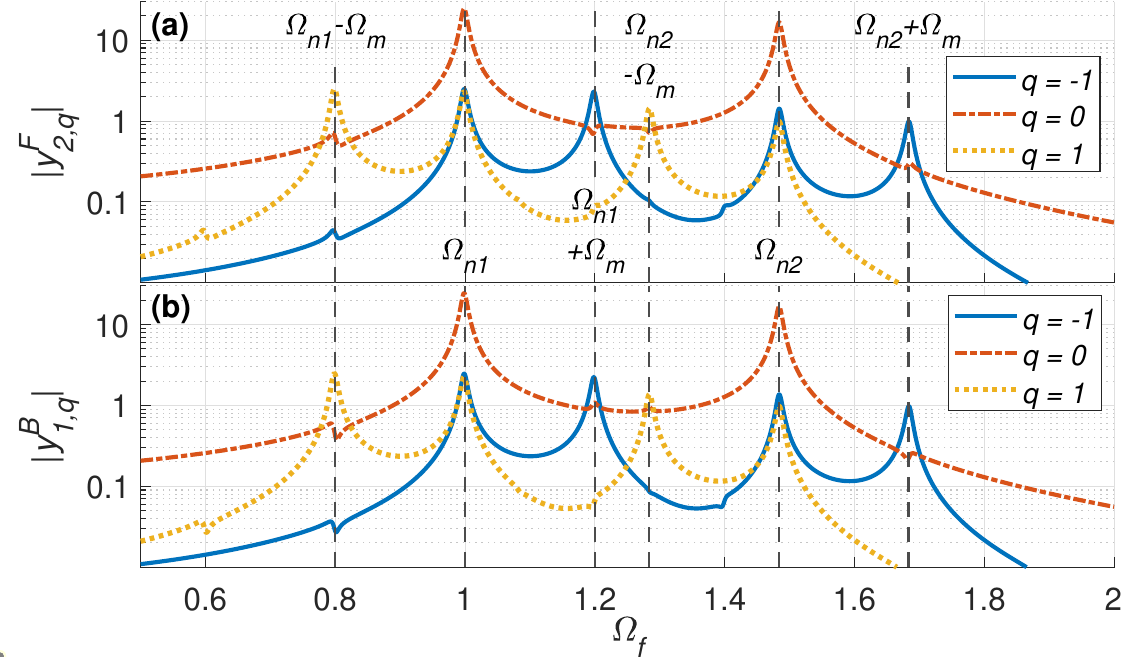}
\caption{\label{fig:Major-Modes1} Plots of the amplitudes of three components ($q\in\{-1,0,1\}$) of (a) \textit{forward} output and (b) \textit{backward} output, and (c) the amplitude difference of each component-pair.}
\end{figure}
\begin{figure}[htb]
\includegraphics[scale=0.45]{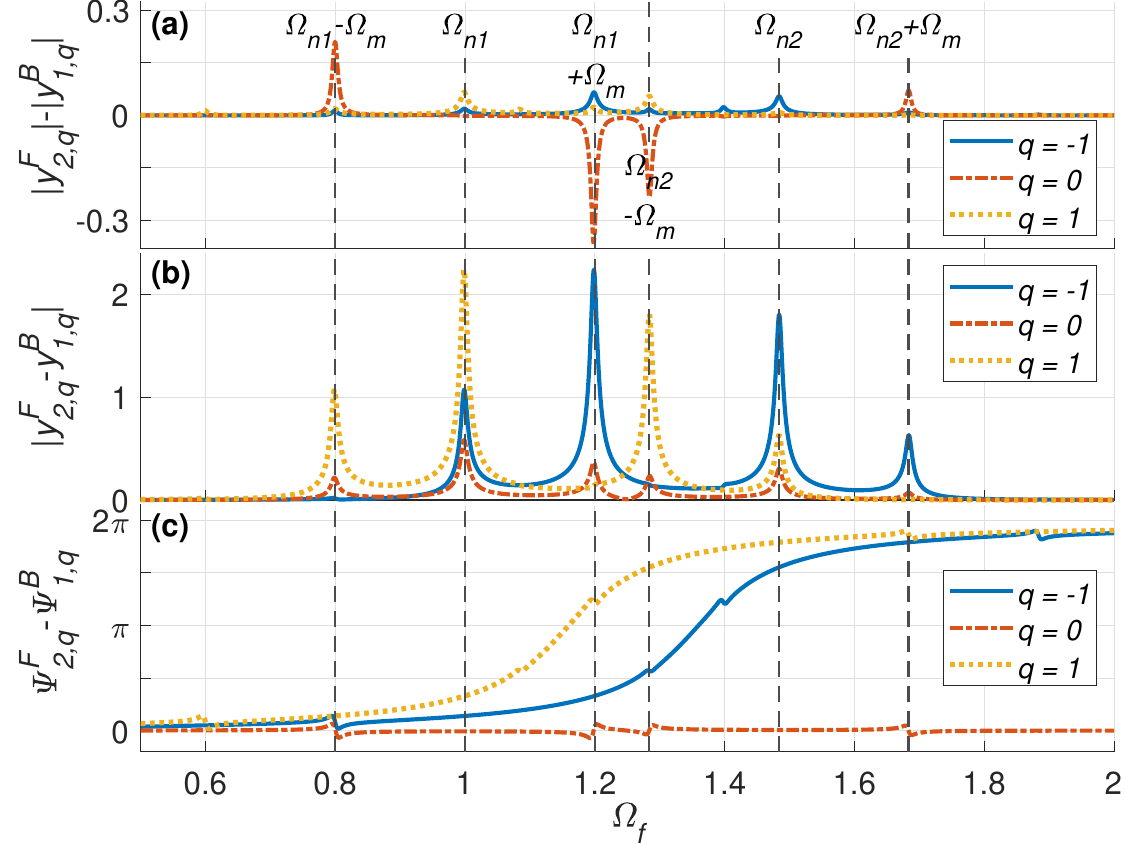}
\caption{\label{fig:Major-Modes2} Plots of (a) the amplitude difference, (b) the phase difference and (c) the contribution of each component-pair to the reciprocity bias.}
\end{figure}
Fig.~\ref{fig:Major-Modes1} shows the amplitudes of the three harmonic components of the response for the \textit{forward} and \textit{backward} configurations; {\it i.e.} the component-pairs.
The component $q=0$ (primary resonance) has the highest amplitudes at $\Omega_{n1}$ and $\Omega_{n2}$, as expected, but it is not a significant contributor at sideband resonances. 
At the upper sideband resonances, $\Omega_{n1}+\Omega_m$ and $\Omega_{n2}+\Omega_m$, the largest component is $q=-1$ for both the \textit{forward} and \textit{backward} configurations. Similarly, the largest component is $q=1$ at the lower sideband resonances, $\Omega_f=\Omega_{n1,2}-\Omega_m$. 

Fig.~\ref{fig:Major-Modes2}(a) shows the difference in the amplitudes of the component-pairs; {\it cf.} Fig.~\ref{fig:Major-Modes1}. As expected, the differences in the amplitudes are too small to account for the reciprocity bias observed in Fig.~\ref{fig:NF-Omg-1}(c). 

Fig.~\ref{fig:Major-Modes2}(b) shows the magnitude of the amplitude difference, $|y_{2,q}^{F}\!-\!y_{1,q}^{B}|$, which accounts for contributions from phase. We see that the component-pairs $q=\pm1$ make the biggest contributions to reciprocity bias. Consistent with Fig.~\ref{fig:Major-Modes1}, the component-pair with $q=1$ contributes most strongly at the lower sideband resonances, while the component-pair with $q=-1$ contributes most strongly at the upper sidebands. Notably, this is in contrast to Fig.~\ref{fig:Major-Modes2}(a), in which the contributions from phase were ignored. 

The role of phase is also observed in the primary component-pair $q=0$: although the difference between the amplitudes is relatively large near the sideband resonances in Fig.~\ref{fig:Major-Modes2}(a), their contribution to reciprocity bias is relatively small in Fig.~\ref{fig:Major-Modes2}(b) where the phase effect is taken into account. 

To complete the picture, Fig.~\ref{fig:Major-Modes2}(c) shows the phase difference for each component-pair. The phase difference in the component-pair $q=0$ agrees with the contrasting behavior observed in panels (a) and (b). The component-pairs $q\pm1$ undergo significant phase changes, which contributes to the reciprocity bias. Note, also, that the three curves in Fig.~\ref{fig:Major-Modes2}(c) never intersect with the horizontal line ($\Psi^F_{2,q}-\Psi^B_{1,q}=0$) at the same $\Omega_f$. This implies that the difference between transmitted phases always contributes to the reciprocity bias.

In summary, we found the transmitted phase to be the main contributor to nonreciprocity in short system with weak modulation. The largest contribution to phase was associated to component-pairs $q=\pm1$.

\subsection{The role of $\bm{\phi}$}
\label{sec:phi}

\begin{figure}[b]
\includegraphics[scale=0.45]{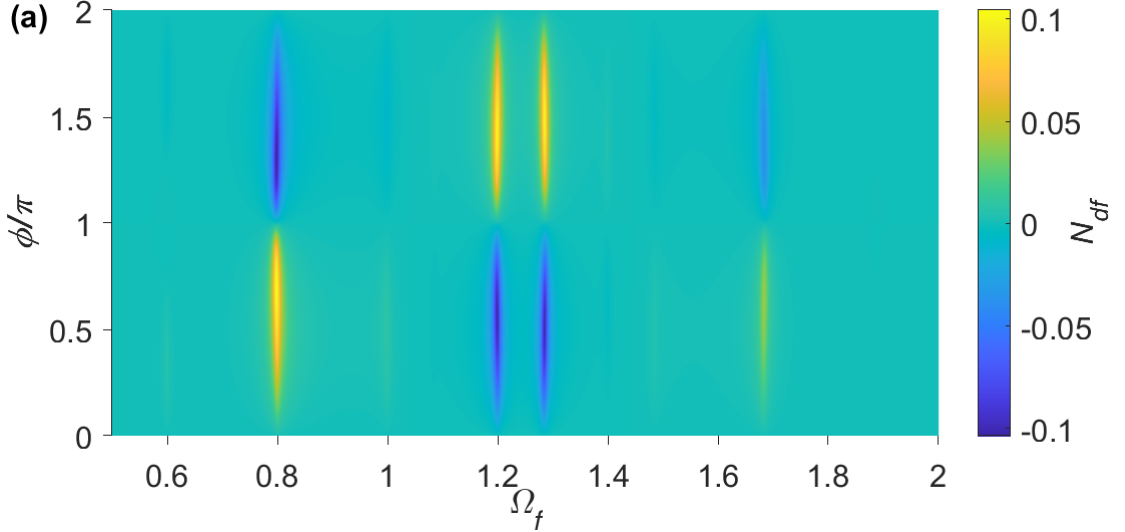}
\includegraphics[scale=0.45]{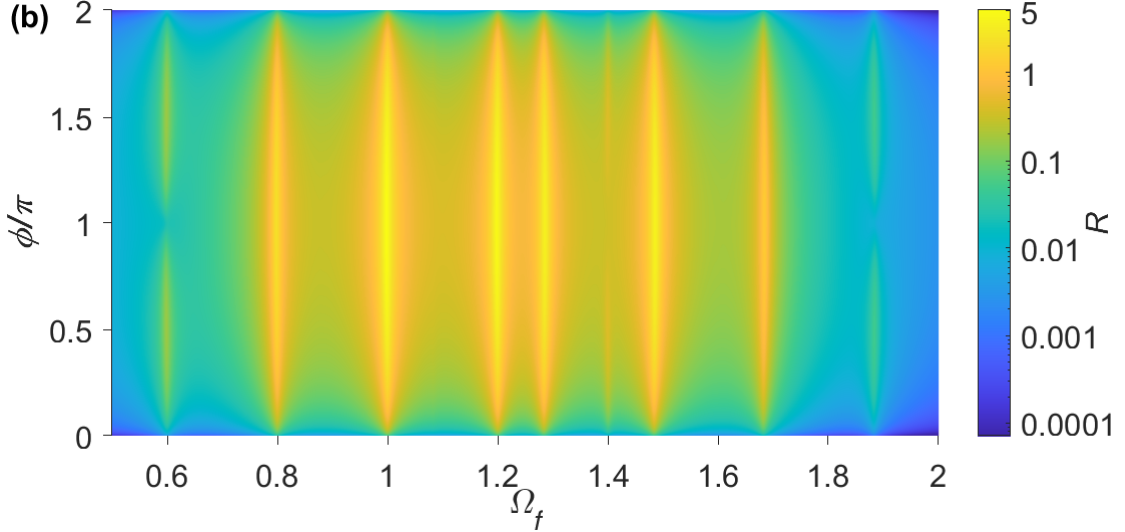}
\caption{\label{fig:surf_weak} Plots of (a) $N_{df}$ and (b) $R$ as functions of $\Omega_f$ and $\phi$.}
\end{figure}

The parameter $\phi$ represents a relative phase shift between the modulations of the two grounding springs in Fig.~\ref{fig:2dof}. This phase shift represents a spatial modulation in the grounding stiffness coefficient of the system. It is the same as the modulation wavenumber in a spatiotemporally modulated system. However, we do not refer to $\phi$ as the modulation wavenumber because the system we study has only two units. Note that the modulation phase, $\phi$, is the only difference between the two oscillators. Thus, it takes on the role of breaking the mirror-symmetry of the system: if $\phi=0$, the response of the system remain reciprocal by virtue of mirror symmetry.

Fig.~\ref{fig:surf_weak} shows the surface plots of $N_{df}$ and $R$ as functions of $\Omega_f$ and $\phi$ for the same parameters used in Fig.~\ref{fig:NF-Omg-1}. We observe that $N_{df}$ changes sign along $\phi=\pi$. This is also a line of symmetry for $R$, implying that $R$ has a local maximum when $\phi=\pi$. We will discuss this behavior in more detail in Section~\ref{phase-nonreciprocity}.

To explain the symmetries observed in Fig.~\ref{fig:surf_weak}, we consider the complex-valued amplitudes $y_{2,q}^F$ and $y_{1,q}^B$. It can be obtained from Eqs.~(\ref{eqC5}) and~(\ref{eqC6}), that:
\begin{equation}
\label{eq5}
y_{2,0}^F(\phi)=y_{1,0}^B(2\pi-\phi),\hspace{.2cm}|y_{2,q}^F(\phi)|=|y_{1,q}^B(2\pi-\phi)|,
\end{equation}
where $q\in[-\mathcal{F},\mathcal{F}]$. The plot of $N_{df}$ is therefore odd-symmetric about the line $(\phi,N_{df})=(\pi,0)$, as seen in Fig.~\ref{fig:surf_weak}(a). Furthermore, $N^F=N^B$ when $\phi=\pi$, regardless of the value of $\Omega_f$.
For the corresponding phases, we have:
\begin{equation}
\label{eq6}
    \Psi^F_{2,q}(\phi)-\Psi^B_{1,q}(\phi)=\Psi^B_{1,q}(2\pi-\phi)-\Psi^F_{2,q}(2\pi-\phi).
\end{equation}
The plot of $R$ is therefore symmetric about the plane $\phi=\pi$, as shown in Fig.~\ref{fig:surf_weak}(b).

The relations in \eqN{eq5} and \eqN{eq6} are valid regardless of the values of all other system parameters; they hold even in systems with more units~\cite{IDETC2023}.
Thus, the odd-symmetry of $N_{df}$ and the symmetry of $R$ persist with the change of system parameters.

The six resonant frequencies of a weakly modulated system correspond to the zeros of the determinant of matrix $\underline{\underline{D}}$ in \eqN{eqC3} with $\zeta=0$ and $\mathcal{F}=1$. This determinant can be expanded as:
\begin{equation}
\label{eqDET-D}
    |\underline{\underline{D}}|=\mathcal{D}_{0}+\epsilon^2\mathcal{D}_{2}+O(\epsilon^4)
\end{equation}
where
\begin{subequations}
\begin{eqnarray}
    \mathcal{D}_{0}=A_{-1}^2A_0^2A_1^2-K_c^2(A_{-1}^2A_0^2+A_{-1}^2A_1^2+A_0^2A_1^2)\nonumber\\
    +K_c^4(A_{-1}^2+A_0^2+A_1^2)-K_c^6\nonumber\hspace{2.3cm}\\
    \mathcal{D}_{2}=2A_0(K_c^2-A_{-1}A_1)(A_{-1}+A_1)\nonumber\hspace{1.9cm}\\
    -2K_c^2(A_{-1}^2+A_1^2-2K_c^2)\cos\phi\nonumber\hspace{2cm}\\
    \epsilon=K_m/2\nonumber\hspace{5.6cm}
\end{eqnarray}
\end{subequations}
and $A_{-1}$, $A_0$ and $A_1$ can be obtained from \eqN{eqCA_j}. For the weakly modulated system studied in this section, we have $\epsilon\leq0.05$. Therefore, $|\underline{\underline{D}}|\approx\mathcal{D}_{0}$ throughout the frequency range considered, $0.5\le\Omega_f\le2$. The six resonant frequencies can be approximated by solving $\mathcal{D}_{0}(\Omega_f)=0$, which gives $\Omega_{n1,2}$ and $\Omega_{n1,2}\pm\Omega_m$. These six frequencies do not depend on the modulation phase, $\phi$. This is why the regions of high amplitude in Fig.~\ref{fig:surf_weak} appear as vertical stripes. We will see in Section~\ref{sec:strong} that this is not true in strongly modulated systems.

\subsection{Overlap of two resonant frequencies}
\label{sec:overlap}
\begin{figure*}
\includegraphics[scale=0.45]{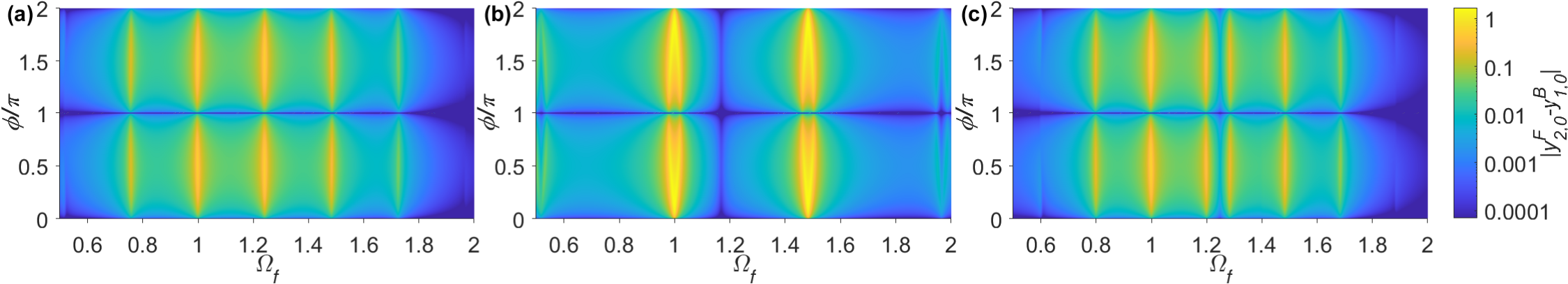}
\includegraphics[scale=0.45]{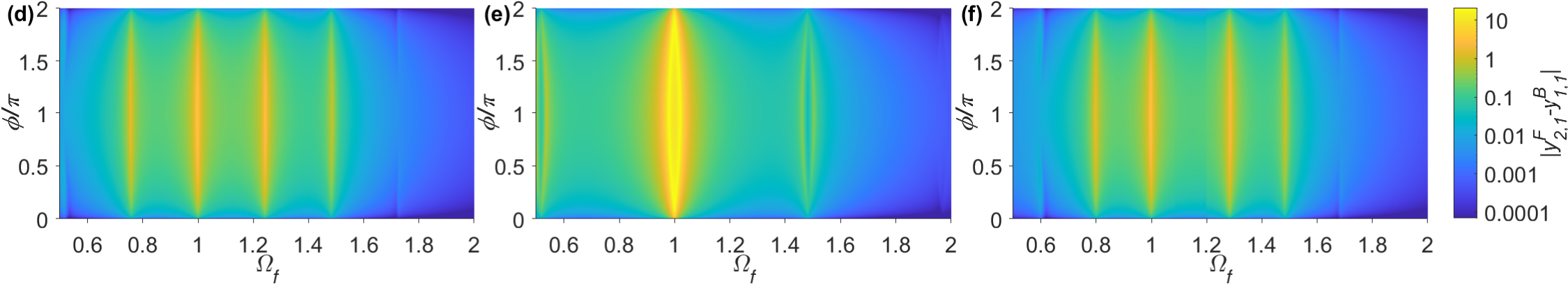}
\includegraphics[scale=0.45]{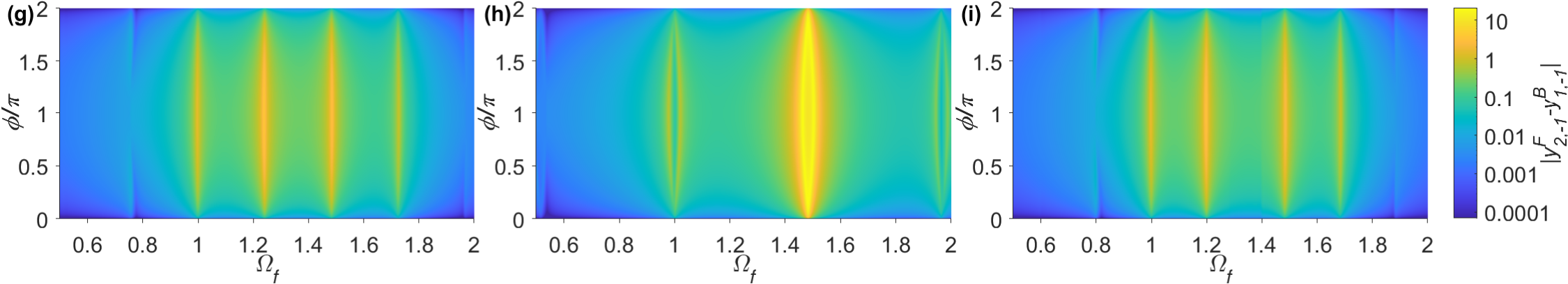}
\caption{\label{fig:overlap1} Plots of $|y^F_{2,q}-y^B_{1,q}|$ for $q\in\{-1,0,1\}$ as functions of $\Omega_f$ and $\phi$ for three systems with different values of $\Omega_m$: (a,d,g) \textit{Case A} with $\Omega_m=(\Omega_{n2}-\Omega_{n1})/2$; (b,e,h) \textit{Case B} with $\Omega_m=\Omega_{n2}-\Omega_{n1}$; (c,f,i) \textit{Case C} with $\Omega_m=0.2$.}
\end{figure*}

The primary ($\Omega_{n1,2}$) and sideband ($\Omega_{n1,2}\pm\Omega_m$) frequencies can be tuned by changing the values of coupling stiffness, $K_c$, and modulation frequency, $\Omega_m$. In this section, we keep $K_c=0.6$ and change $\Omega_m$ to investigate the influence of frequency overlaps.

Fig.~\ref{fig:overlap1} shows the variation of $|y^F_{2,q}-y^B_{1,q}|$ as functions of $\Omega_f$ and $\phi$ for component-pairs with $q=0$ (first row), $q=1$ (second row) and $q=-1$ (third row). We consider two scenarios with resonant frequency overlaps: (i) \textit{Case A}: a system with $\Omega_m=(\Omega_{n2}-\Omega_{n1})/2$, where two sidebands overlap (left column in Fig.~\ref{fig:overlap1}); (ii) \textit{Case B}: a system with $\Omega_m=\Omega_{n2}-\Omega_{n1}$, where each primary band overlaps with a sideband (middle column in Fig.~\ref{fig:overlap1}). A third scenario, \textit{Case C}, is shown in the right column of Fig.~\ref{fig:overlap1}, where there is no frequency overlap (the same as Fig.~\ref{fig:surf_weak}). 
Except for $\Omega_m$, all system parameters are the same in these three scenarios. The same logarithmic scale is used in each row.

We observe in Fig.~\ref{fig:overlap1} that the magnitude of $|y^F_{2,q}-y^B_{1,q}|$ in \textit{Case B} is significantly higher than those in \textit{Case A} and \textit{Case C} for all component pairs, $q=-1,0,1$. Accordingly, the magnitude of reciprocity bias in \textit{Case B} is the highest among the three (not shown). We also observe that the component-pairs $q=\pm1$ have a more significant contribution to reciprocity bias than the component-pair $q=0$. 
We observe in the top row of Fig.~\ref{fig:overlap1} that $|y^F_{2,0}-y^B_{1,0}|=0$ when $\phi=\pi$, as predited by~\eqN{eq5}.

In \textit{Case B} (middle column in Fig.~\ref{fig:overlap1}), we observe that the regions of high amplitude no longer appear as vertical stripes, as they do in \textit{Case A} and \textit{Case C}. This means that the resonant frequencies have a weak dependence on the modulation phase, $\phi$. This happens because $\mathcal{D}_{0}$ and $\epsilon^2\mathcal{D}_{2}$ in~\eqN{eqDET-D} have the same order of magnitude when two primary bands overlap ($\Omega_m=\Omega_{n2}-\Omega_{n1}$).

\section{Vibration transmission in strongly modulated systems}
\label{sec:strong}
Strong modulations ($K_m>0.1$) bring about different vibration characteristics in spatiotemporally modulated systems. We investigate some of these characteristics in this section. 

Note that increasing the modulation amplitude can result in parametric instabilities, which lead to unbounded response~\cite{MathieuKovacic}. We have computed the stability bounds for our system, and further ensured that all the results presented in this work are stable and remain bounded by direct numerical integration of the governing equations. However, a detailed analysis of parametric instabilities falls outside the scope of the current work and will be presented separately elsewhere. 

\subsection{Steady-state response}
\begin{figure}[b]
\includegraphics[scale=0.45]{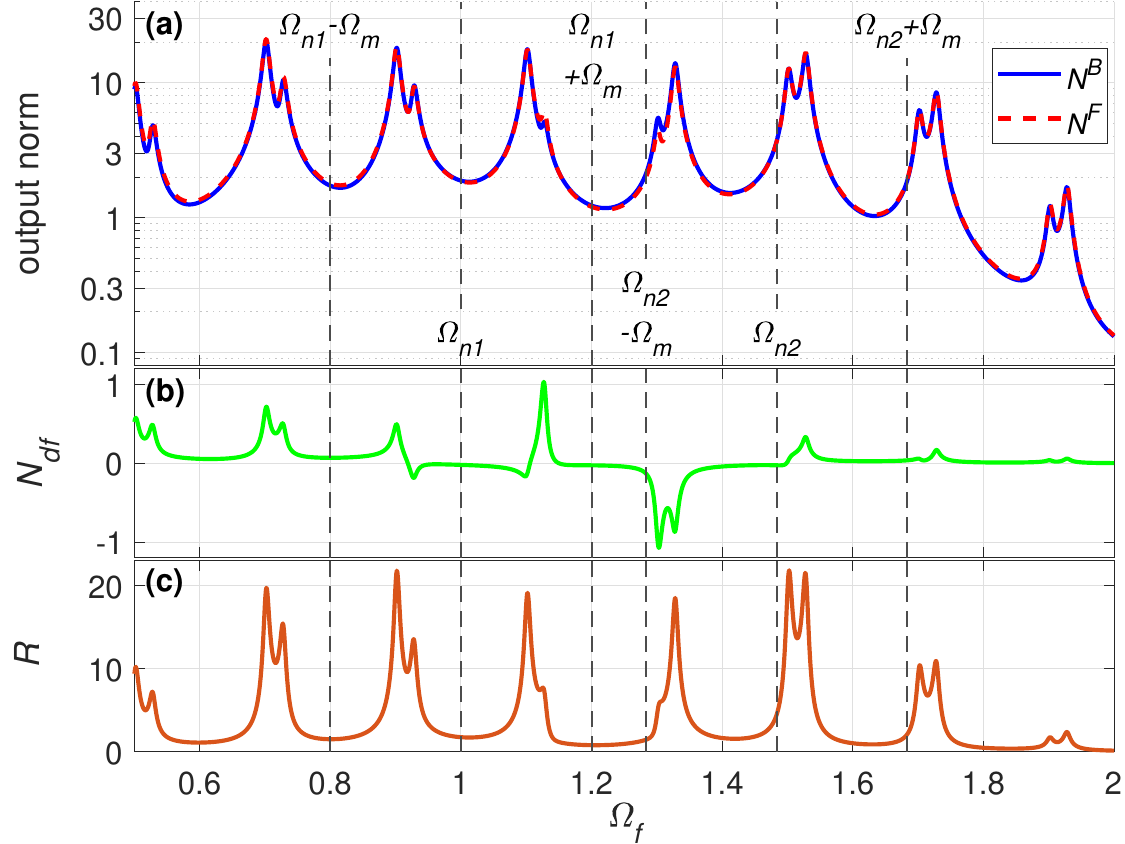}
\caption{\label{fig:LargeKm-Freq} Plots of (a) output norms, (b) difference between output norms and (c) reciprocity bias as functions of forcing frequency.}
\end{figure}

Fig.~\ref{fig:LargeKm-Freq} shows the response of a strongly modulated system with $K_m=0.8$ in the \emph{forward} and \emph{backward} configurations; {\it cf.} Fig.~\ref{fig:NF-Omg-1}. All other system parameters are the same as those used in Section~\ref{sec:sidebands}. 

The number and frequencies of resonance peaks in Fig.~\ref{fig:LargeKm-Freq}(a) are very different from what we observed in weakly modulated systems. The peaks no longer appear at $\Omega_{n1}$, $\Omega_{n2}$, $\Omega_{n1}\pm\Omega_m$ and $\Omega_{n2}\pm\Omega_m$. Sideband resonances are no longer limited to $q\in\{-1,0,1\}$ because the amplitudes of higher-order sidebands ($q>1$) do not diminish as significantly. And it is difficult to distinguish primary and sideband resonances by their relative peak amplitudes. Even though the values of the peak amplitudes are similar to those in the weakly modulated system, there is clearly more energy in the strongly modulated system; compare the areas under the frequency response functions in Figs.~\ref{fig:LargeKm-Freq}(a) and~\ref{fig:NF-Omg-1}(a).

Despite these key differences between weakly and strongly modulated systems, the transmitted phase remains a signficant contributor to the reciprocity bias. Figs.~\ref{fig:LargeKm-Freq}(b) and \ref{fig:LargeKm-Freq}(c) show the difference between transmitted energies, $N_{df}$, and reciprocity bias, $R$, respectively. We observe that although the difference in transmitted energies is relatively small, reciprocity bias is very large in comparison. This indicates the important role of phase in breaking reciprocity. 
\begin{figure}[tb]
\includegraphics[scale=0.45]{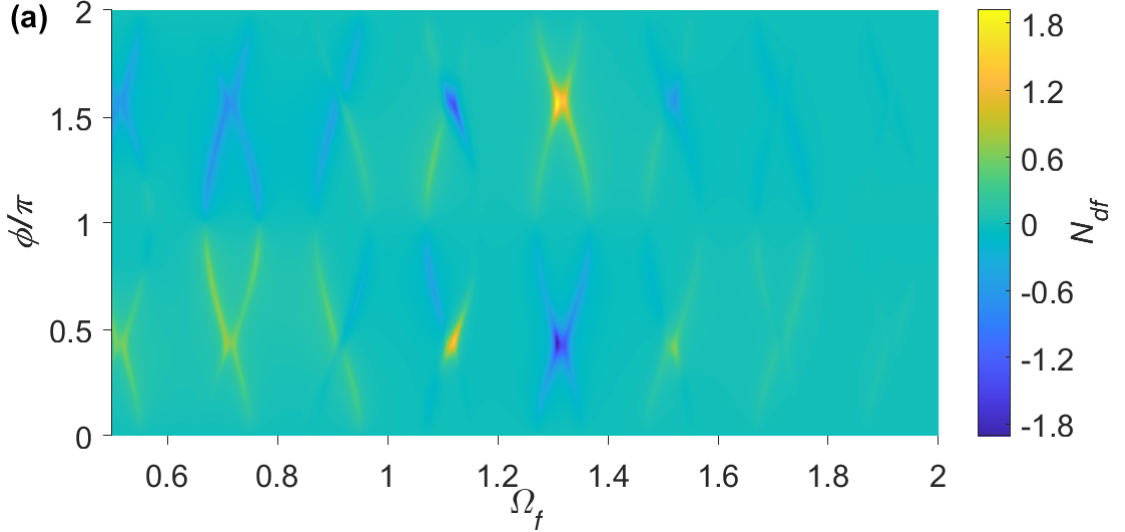}
\includegraphics[scale=0.45]{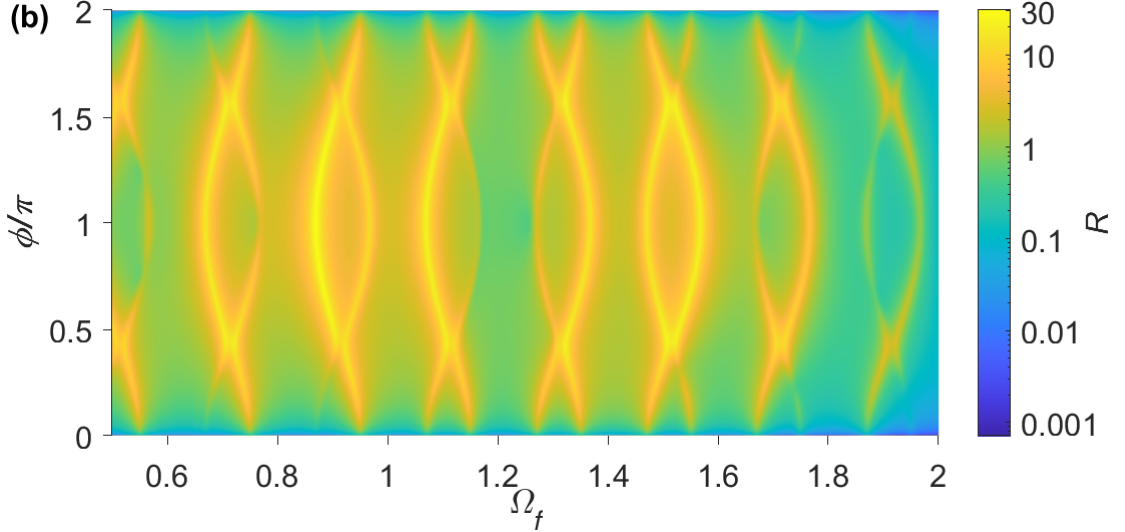}
\caption{\label{fig:surf_strong} Plots of (a) $N_{df}$ and (b) $R$ as functions of $\Omega_f$ and $\phi$.}
\end{figure}

Fig.~\ref{fig:surf_strong} shows the variation of $N_{df}$ and $R$ as functions of $\Omega_f$ and $\phi$. The symmetry properties discussed in Section~\ref{sec:phi} still hold because they do not depend on the strength of modulation. Most notably, we observe that the peak frequencies depend on the modulation phase, $\phi$, in stark contrast to weakly modulated systems; {\it cf.} Fig.~\ref{fig:surf_weak}. We explore this phenomenon in the next section.

\subsection{Resonant frequencies}
\begin{figure}[htb]
\includegraphics[scale=0.45]{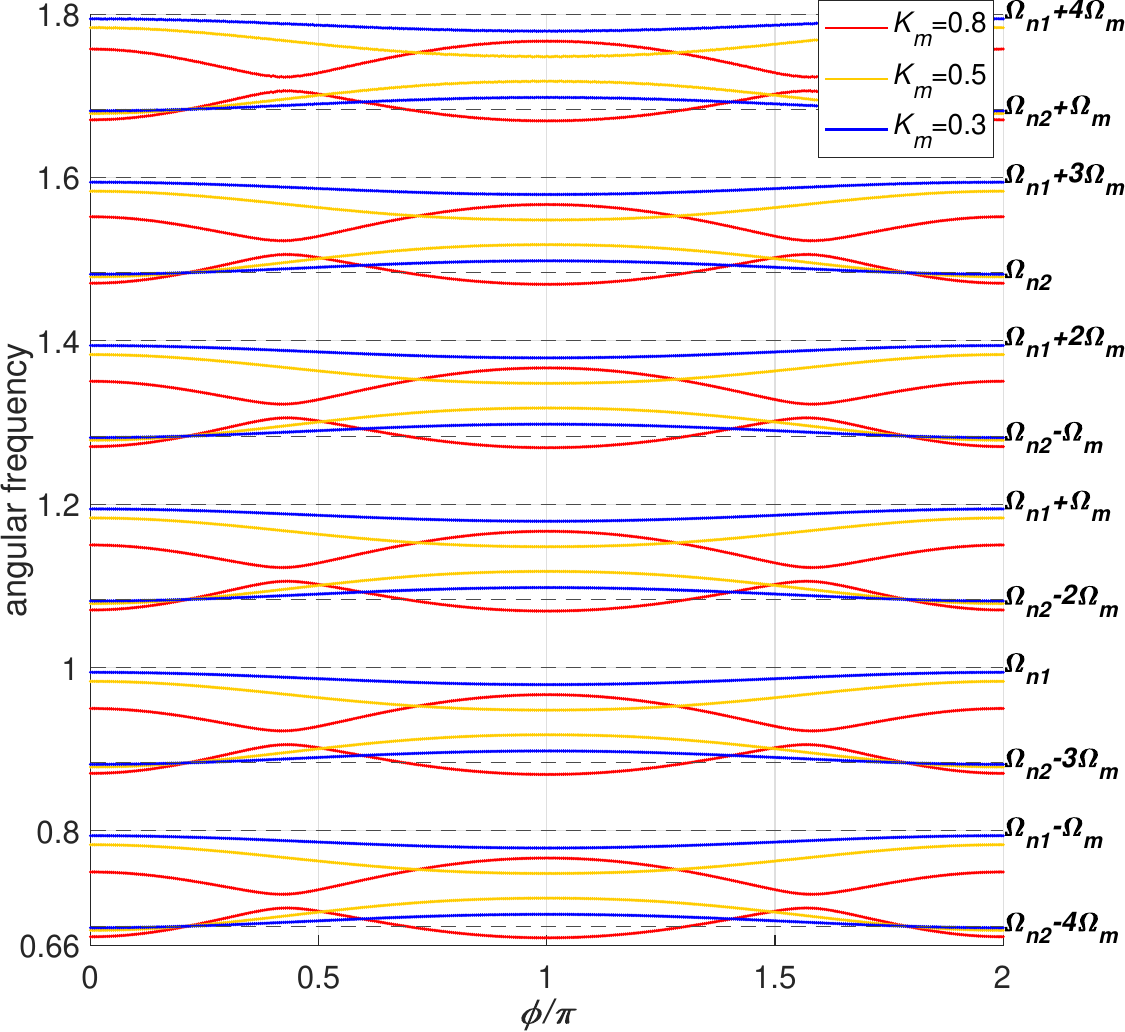}
\caption{\label{fig:Km-Freq} Resonant frequencies of a strongly modulated system as a function of $\phi$.}
\end{figure}

\begin{figure}[b]
\includegraphics[scale=0.45]{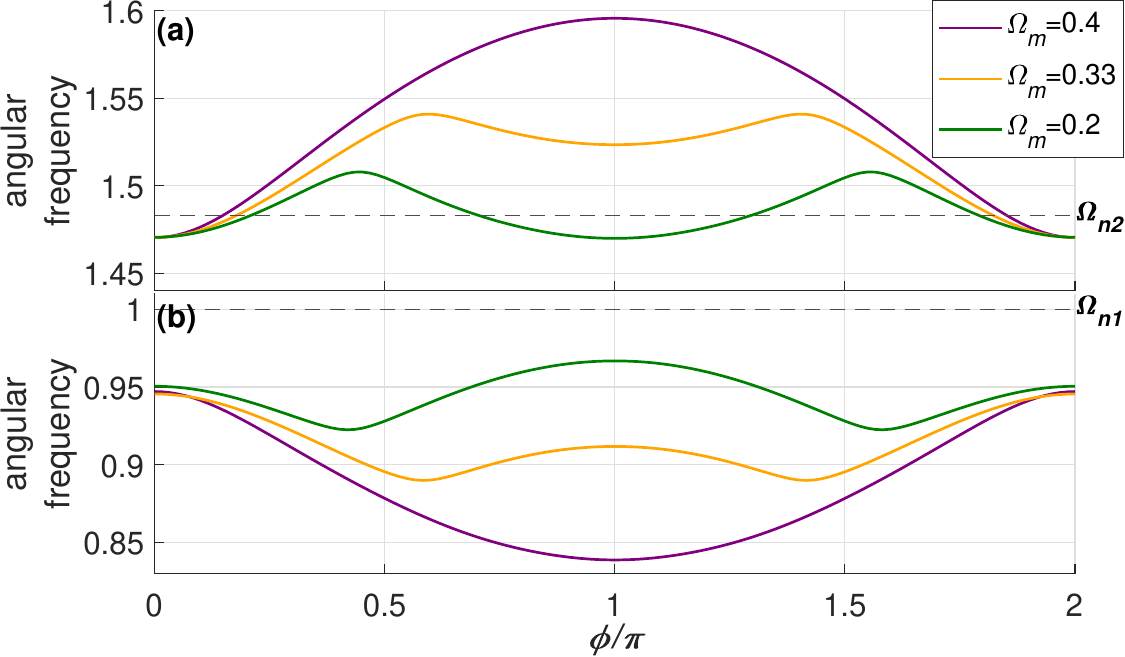}
\caption{\label{fig:Wm-Freq} resonant frequencies nearby (a) $\Omega_{n1}$ and (b) $\Omega_{n2}$ as functions of $\phi$.}
\end{figure}

We calculate the resonant frequencies of the modulated system based on the formulation developed in Appendix~\ref{appendix:avg}. The resonant frequencies of the systems are the zeros of the determinant of the matrix $\underline{\underline{D}}$ in \eqN{eqC3} when $\zeta=0$; {\it i.e.}, $|\underline{\underline{D}}|=0$. The response amplitude in this case becomes infinite, as expected.

In weakly modulated systems (small $K_m$), the first super diagonal and the first subdiagonal of $\underline{\underline{D}}$ are negligible compared to its main diagonal. Because the modulation parameters $K_m$ and $\phi$ do not appear on the main diagonal, they have little influence on the solutions of $|\underline{\underline{D}}(\Omega_f)|=0$. Therefore, the resonant frequencies of the weakly modulated system are mainly determined by $K_c$ and $\Omega_m$; see \eqN{eqDET-D}. 
In strongly modulated systems, the entries on the first super diagonal and the first subdiagonal of $\underline{\underline{D}}$ are no longer negligible. Therefore, the resonant frequencies of the system depend on the values of $K_m$ and $\phi$ too.

Fig.~\ref{fig:Km-Freq} shows the natural frequencies of the strongly modulated systems as a function of $\phi$ for $K_m\in\{0.3,0.5,0.8\}$, $K_c=0.6$ and $\Omega_m=0.2$. The horizontal dashed lines denote $\Omega_{n1}\pm q\Omega_m$ and $\Omega_{n2}\pm q\Omega_m$ with $q=0,\cdots,4$.
We observe that all the loci in Fig.~\ref{fig:Km-Freq} have local maxima at $\phi\in\{0,\pi,2\pi\}$, a property that stems from the symmetries of the cosine function in the modulation term. 
As $K_m$ increases to 0.3 and 0.5, the deviations in natural frequencies are largest near $\phi=\pi$, and decrease monotonically away from this point. The variations in the natural frequencies are no longer monotonic for higher strengths of modulation, as observed for $K_m=0.8$. We also observe avoided crossing between adjacent branches; see the loci of $\Omega_{n2}$ to $\Omega_{n1}\!+\!3\Omega_m$ for an example of this.

Fig.~\ref{fig:Wm-Freq} shows the variation of the primary natural frequencies, $\Omega_{n1}$ and $\Omega_{n2}$, as a function of modulation phase for $\Omega_m\in\{0.2,0.33,0.4\}$. We observe that at larger values of the modulation frequency, the range of variation in $\Omega_{n1}$ and $\Omega_{n2}$ is larger as well.

\section{Phase Nonreciprocity}
\label{phase-nonreciprocity}
Looking back at \eqN{eq:norms0}, if $R=0$ then it is obvious that $N^F=N^B$; {\it i.e.}, if the response is reciprocal then the transmitted energies in the {\it forward} and {\it backward} directions are identical. However, $N^F=N^B$ cannot guarantee a reciprocal response. This means that it is possible to have nonreciprocal response ($R\ne0$) that is accompanied by equal energies transmitted in opposite directions ($N^F=N^B$). We refer to this scenario as {\it phase nonreciprocity} because a difference in the transmitted phases is the sole contributor to nonreciprocity. Phase nonreciprocity in vibration transmission has been reported in time-invariant nonlinear systems~\cite{nonlinear_2DoF,nonlinear_phase}. 
\begin{figure}[hbt]
\includegraphics[scale=0.45]{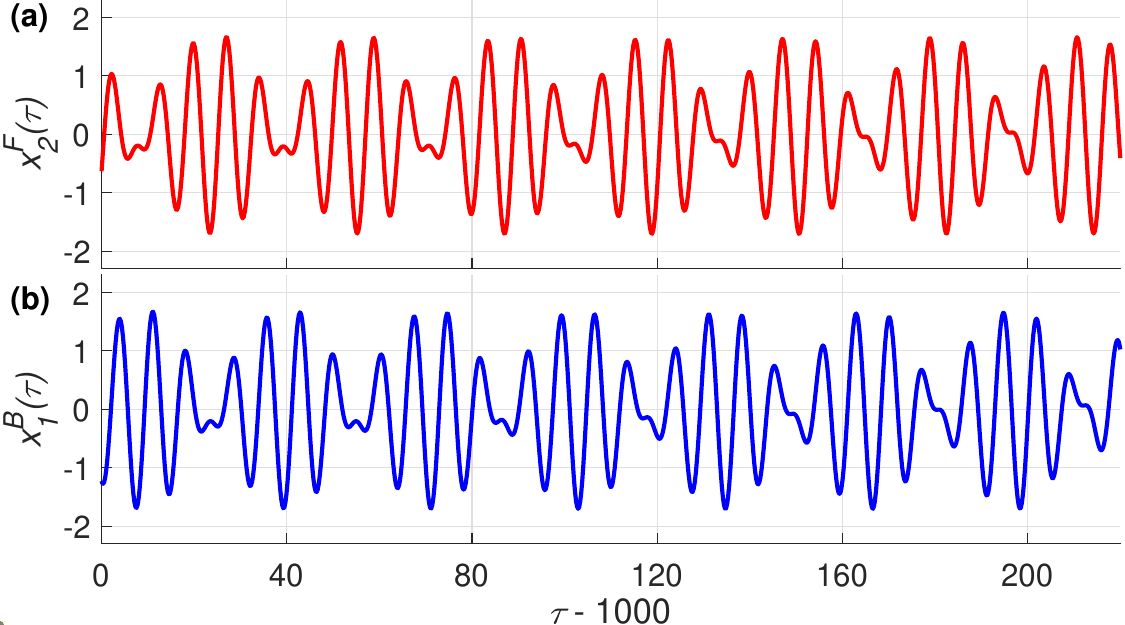}
\includegraphics[scale=0.45]{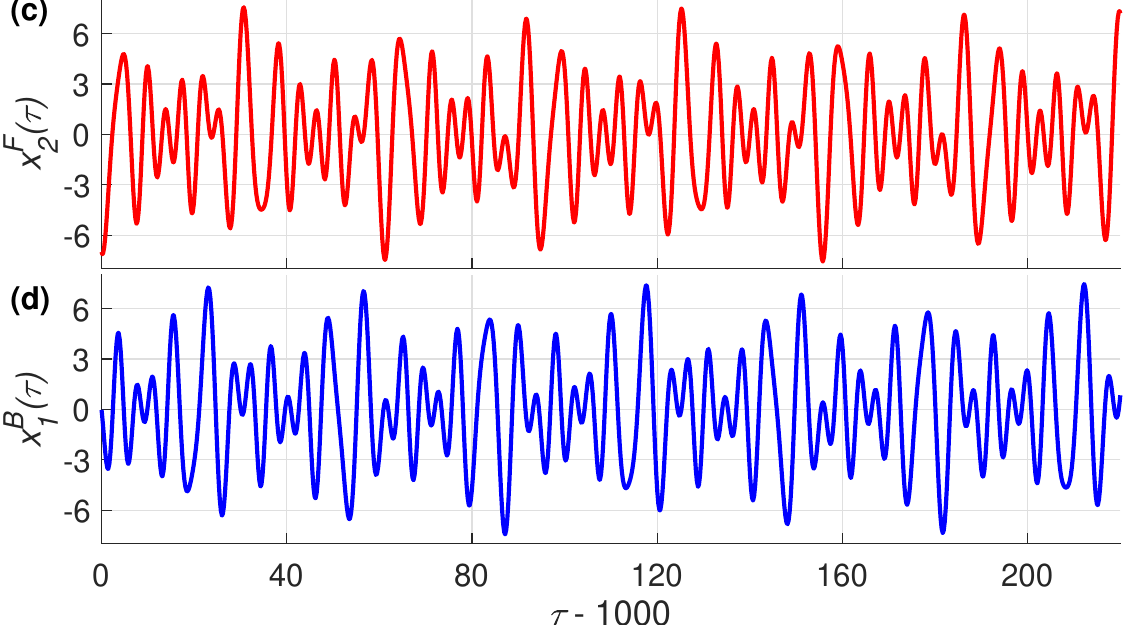}
\caption{\label{fig:phase-nonrecp-time} Plots of \textit{forward} and \textit{backward} outputs. Common parameters in these two examples: $K_c=0.6$, $\zeta=0.005$, $\Omega_m=0.2$ and $P=1$. (a,b) $\phi=\pi$, $\Omega_f=0.79$ and $K_m=0.1$; (c,d) $\phi=0.75\pi$, $\Omega_f=0.93$ and $K_m=0.8$.}
\end{figure}

Due to the odd-symmetry of $N_{df}$ about the line $(\phi,N_{df})=(\pi,0)$, a trivial case of phase nonreciprocity occurs when $\phi=\pi$, regardless of the values of other system parameters. This is because the matrix $\underline{\underline{D}}$ in \eqN{eqC3} becomes symmetric for $\phi=\pi$. Therefore, the amplitudes of each pair of harmonic components, $|y^F_{2,q}|$ and $|y^B_{1,q}|$, become equal for $q=0,\pm1,\pm2,\cdots$. Meanwhile, $\Psi^F_{2,q}-\Psi^B_{1,q}=\pi$ if $q$ is an odd number; $\Psi^F_{2,q}-\Psi^B_{1,q}=0$ if $q$ is an even number.

If $\phi\neq\pi$, there exist combinations of $\Omega_f$ and $\phi$ which can lead to $N_{df}=0$. Because $R>0$ throughout the ranges of $\Omega_f$ and $\phi$ considered, the response at these combinations of $\Omega_f$ and $\phi$ is therefore phase nonreciprocal.
Fig.~\ref{fig:phase-nonrecp-time} shows the outputs in the time domain for two examples of phase nonreciprocity with $\phi=\pi$ and $\phi\neq\pi$.
While it is obvious that $x^F_2(\tau)\neq x^B_1(\tau)$, the transmitted vibrations have the same amount of energy, $N^F=N^B$.

For the non-trivial case of $\phi\ne\pi$, a more stringent requirement than equal transmitted energies ($N^F=N^B$) is to have nonreciprocal transmission with the same waveform. We were only able to find parameters that lead to this scenario in systems with more than two degrees of freedom~\cite{JiudaASAphase}. The methodology involved for these calculations falls beyond the scope of this paper. We postpone this discussion to a future publication.

\section{Conclusions}
We investigated nonreciprocal vibration transmission in a system of coupled mechanical oscillators subject to spatiotemporal stiffness modulations. The temporal modulation appeared as harmonic modulation of the grounding stiffness of each oscillator. The phase difference between the two temporal modulations ($\phi$) acts as the spatial modulation, equivalent to the modulation wavenumber in a longer system. The modulation phase, $\phi$, acts as the symmetry-breaking parameter that is necessary to break reciprocity. We used the averaging method to develop an analytical framework to obtain the steady-state quasi-periodic response of the system to harmonic external excitation. These results were validated against direct numerical simulation of the response of the system for both weak and strong modulations. 

We found the response to be nonreciprocal when $\phi\ne0$, as expected. However, the transmitted energies in the forward and backward configurations were similar in most cases, meaning that the difference between the transmitted phases is the main contributor to breaking reciprocity in short systems. This was the case for both weak and strong modulations. 

In weakly modulated systems, we found only one pair of sideband resonances to be sufficient to capture the response of the system accurately. The pairs of harmonic components of the response (primary and sideband) contribute differently to the reciprocity bias. We found that an overlap of two primary frequencies results in stronger nonreciprocity than an overlap of a primary and sideband resonance.

Increasing the strength of modulations significantly increases the reciprocity bias because there is more energy provided to the system. Increasing the modulation amplitude also makes the resonance frequencies dependent on the modulation phase and amplitude. The frequency contents of the response of strongly modulated systems is richer due to contributions from additional (higher-order) sideband frequencies.

We found two types of nonreciprocal response in which equal amounts of energy is transmitted in the forward and backward configurations. In one case, the two output displacements are distinguished by just a phase shift, whereas in the other the two waveforms are different while maintaining the same energy. This feature will be addressed in detail in the near future, along with an analysis of parametric instabilities for strongly modulated systems. The analytical framework developed here paves the way for these studies.

\begin{acknowledgments}
We acknowledge financial support from the Natural Sciences and Engineering Research Council of Canada through the Discovery Grant program. J.W. acknowledges additional support from Concordia University and from {\it Centre de Recherches Mathématique}, Quebec.
\end{acknowledgments}

\appendix

\section{Non-dimensionalization}
\label{appendix:nondimensionalization}
The equations of motion which govern the 2-DoF modulated system in Fig.~\ref{fig:2dof} read:
\begin{subequations}
\label{eqA1}
\begin{eqnarray}
m\frac{d^2u}{dt^2}\!+\!c\frac{du}{dt}\!+\!k_1u\!+\!k_c(u\!-\!v)\!=\!F_1\cos(\omega_ft),\label{eq:A1a} \\
m\frac{d^2v}{dt^2}\!+\!c\frac{dv}{dt}\!+\!k_1^\prime v\!+\!k_c(v\!-\!u)\!=\!F_2\cos(\omega_ft), \label{eq:A1b}
\end{eqnarray}
\end{subequations}
where $k_1=k_{g,DC}+k_{g,AC}\cos(\omega_mt)$ and $k_1^\prime=k_{g,DC}+k_{g,AC}\cos(\omega_mt-\phi)$. 
We define $\tau=\omega_0t$, where $\omega_0=\sqrt{k_{g,DC}/m}$.
Therefore, $d/dt=\omega_0 d/d\tau$, $d^2/dt^2=\omega_0^2 d^2/d\tau^2$.

To non-dimensionalize, we define $\zeta=c/(2m\omega_0)$, $\Omega_m=\omega_m/\omega_0$, $\Omega_f=\omega_f/\omega_0$,  $K_c=k_c/k_{g,DC}$, $K_m=k_{g,AC}/k_{g,DC}$, $P_1=F_1/(ak_{g,DC})$, $P_2=F_2/(ak_{g,DC})$, $x_1=u/a$ and $x_2=v/a$, where $a$ is a representative length.
After substituting them into \eqN{eqA1}, we obtain:
\begin{subequations}
\label{eqA2}
\begin{eqnarray}
ma\omega_0^2\ddot{x}_1\!+\!2\zeta ma\omega_0^2\dot{x}_1\!+\!k_{g,DC}ax_1\left[1\!+\!K_m\cos(\Omega_m\tau)\right]\nonumber \\
+K_ck_{g,DC}a(x_1\!-\!x_2)\!=\!P_1ak_{g,DC}\cos(\Omega_f\tau),\nonumber \\
\label{eq:A2a} \\
ma\omega_0^2\ddot{x}_2\!+\!2\zeta ma\omega_0^2\dot{x}_2\!+\!k_{g,DC}ax_2\left[1\!+\!K_m\cos(\Omega_m\tau\!-\!\phi)\right]\nonumber \\
+K_ck_{g,DC}a(x_2\!-\!x_1)\!=\!P_2ak_{g,DC}\cos(\Omega_f\tau),\nonumber \\
\label{eq:A2b}
\end{eqnarray}
\end{subequations}
where $\ddot{x}$ and $\dot{x}$ represent $\frac{d^2x}{d\tau^2}$ and $\frac{dx}{d\tau}$, respectively.
\eqN{eqA2} can be further simplified as:
\begin{subequations}
\label{eqA3}
\begin{eqnarray}
\ddot{x}_1+2\zeta\dot{x}_1+[1+K_m\cos(\Omega_m\tau)]x_1\hspace{1.75cm}\nonumber\\
\hspace{2.1cm}+K_c(x_1-x_2)=P_1\cos(\Omega_f\tau),\label{eq:A3a}\\
\ddot{x}_2+2\zeta\dot{x}_2+[1+K_m\cos(\Omega_m\tau-\phi)]x_2\hspace{1.1cm}\nonumber\\
+K_c(x_2-x_1)=P_2\cos(\Omega_f\tau).\hspace{.06cm}\label{eq:A3b}
\end{eqnarray}
\end{subequations}

In this paper, calculations and analysis of the response of the 2-DoF modulated system are all based on \eqN{eqA3}, which is the same as \eqN{eq1}.

\section{Frequency contents of the outputs}
\label{appendix:fft}
\begin{figure*}[hbt]
\includegraphics[scale=0.45]{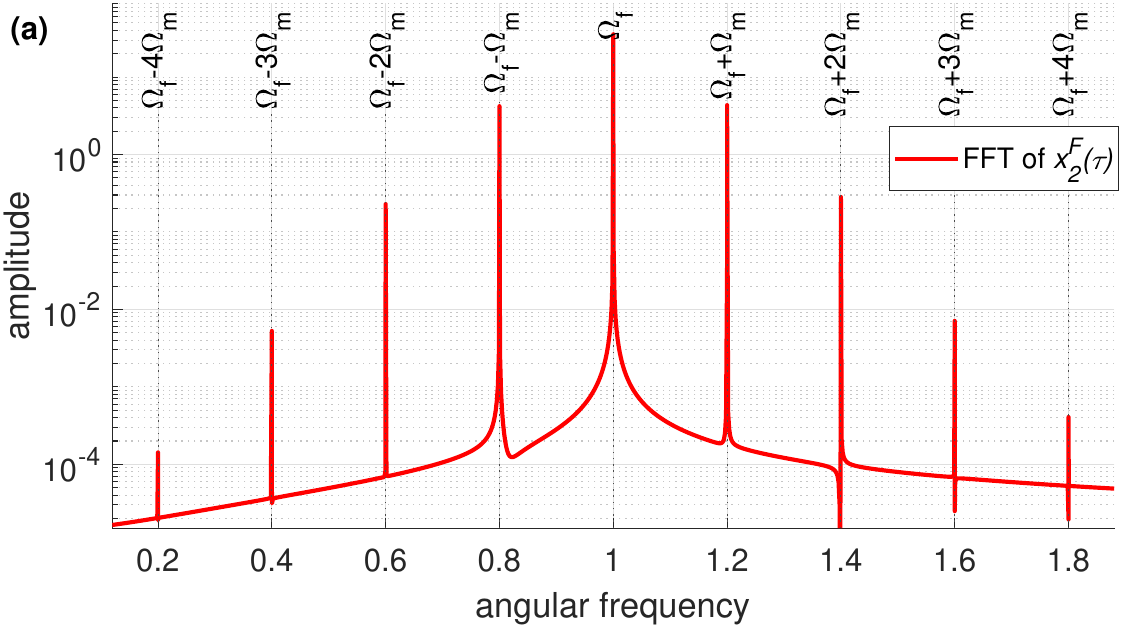}
\includegraphics[scale=0.45]{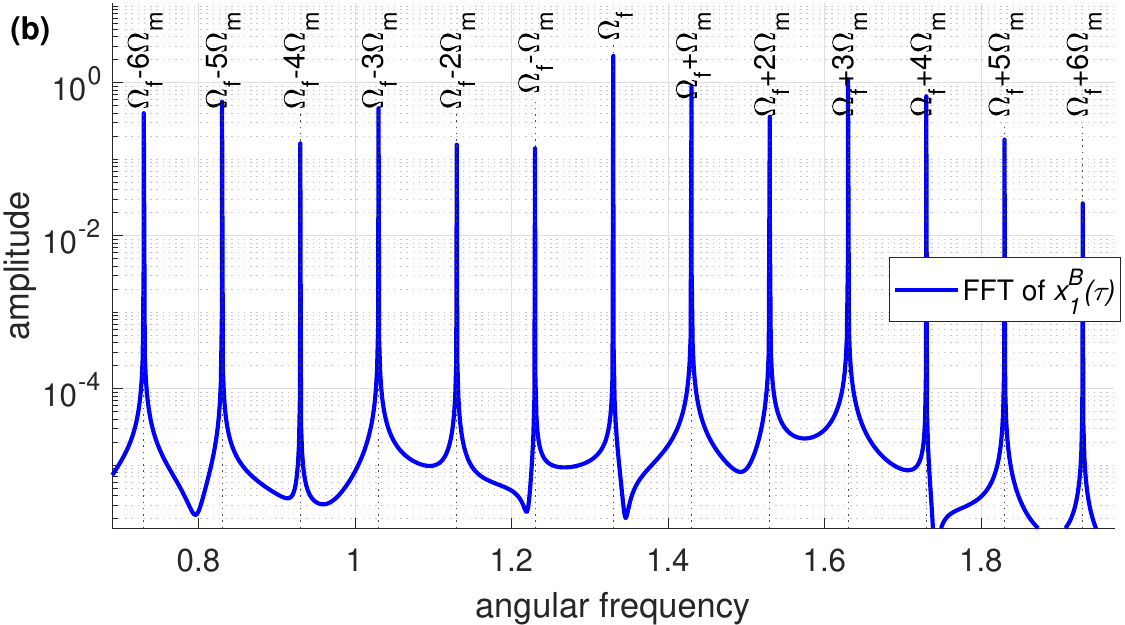}
\caption{\label{fig:FFT-forced} Frequency spectrum of the steady-state response. Parameters in these examples: (a) $K_c=0.6$, $K_m=0.1$, $\zeta=0.005$, $\Omega_m=0.2$, $\phi=0.5\pi$, $P=1$ and $\Omega_f=1$, in \textit{forward} configuration; (b) $K_c=0.7$, $K_m=0.6$, $\zeta=0.005$, $\Omega_m=0.1$, $\phi=0.3\pi$, $P=1$ and $\Omega_f=1.33$, in \textit{backward} configuration.}
\end{figure*}

Due to the simultaneous presence of external and parametric excitations of incommensurate frequencies, it is not straightforward to guess the frequency spectrum of the steady-state output of modulated systems. Here, we use the Runge-Kutta method to obtain the transient response of the modulated system numerically. The output displacement of the system is then recorded after the steady state is reached. We do this for the \textit{forward} configuration with weak modulation and the \textit{backward} configuration with strong modulation. We then obtain the Fast Fourier Transform (FFT) of the steady-state outputs, shown in Fig.~\ref{fig:FFT-forced}.

We observe in Fig.~\ref{fig:FFT-forced} that all the response is dominated by the frequencies $\Omega_f+q\Omega_m$ where $q\in\{\cdots,-2,-1,0,1,2,\cdots\}$. In the case of weak modulations, Fig.~\ref{fig:FFT-forced}(a), the magnitudes (heights) of the peaks decrease rapidly as the frequency moves away from $\Omega_f$ (notice the logarithmic scale for the amplitude). For a strongly modulated system, however, the height of a peak is not directly related to its distance to $\Omega_f$, as shown in Fig.~\ref{fig:FFT-forced}(b).

We conclude for both weakly and strongly modulated systems, that the output can be reasonably approximated using a truncated harmonic expansion with a set of frequencies determined by $\Omega_f$ and $\Omega_m$. This leads to the expressions used in \eqN{eq3}.

\section{Application of the averaging method}
\label{appendix:avg}
We use the system in the \textit{forward} configuration to show the application of the averaging method to approximate the steady-state response of the 2-DoF modulated system. The same procedure can be used for the \textit{backward} configuration.

We start by substituting \eqN{eq3}, the complex Fourier series of the steady-state response,  into \eqN{eq1}, the equations of motion of the system. We use Euler's formula to rewrite the harmonic modulation terms within \eqN{eq1} in the complex exponential form. After algebraic simplifications, we arrive at the following equations:
\begin{widetext}
\begin{subequations}
\label{eqC1}
\begin{eqnarray}
\sum^{\infty}_{q=-\infty} \left[1\!+\!K_c\!-\!(\Omega_f\!+\!q\Omega_m)^{2}\!+\!i2\zeta(\Omega_f\!+\!q\Omega_m)\right]y_{1,q}^{F}e^{iq\Omega_m\tau}\!-\!K_c\!\!\sum^{\infty}_{q=-\infty} y_{2,q}^{F}e^{iq\Omega_m\tau}\hspace{0.9cm}\nonumber\\
+\frac{K_m}{2}\!\!\sum^{\infty}_{q=-\infty} y_{1,q}^{F}e^{i(q+1)\Omega_m\tau}\!+\!\frac{K_m}{2}\!\!\sum^{\infty}_{q=-\infty} y_{1,q}^{F}e^{i(q-1)\Omega_m\tau}\!=\!\frac{P}{2}\;, \label{eqC1a}\\
\sum^{\infty}_{q=-\infty} \left[1\!+\!K_c\!-\!(\Omega_f\!+\!q\Omega_m)^{2}\!+\!i2\zeta(\Omega_f\!+\!q\Omega_m)\right]y_{2,q}^{F}e^{iq\Omega_m\tau}\!-\!K_c\!\!\sum^{\infty}_{q=-\infty} y_{1,q}^{F}e^{iq\Omega_m\tau}\hspace{0.9cm}\nonumber\\
+\frac{K_m}{2}e^{-i\phi}\!\!\sum^{\infty}_{q=-\infty} y_{2,q}^{F}e^{i(q+1)\Omega_m\tau}\!+\!\frac{K_m}{2}e^{i\phi}\!\!\sum^{\infty}_{q=-\infty} y_{2,q}^{F}e^{i(q-1)\Omega_m\tau}\!=\!0
\;.\hspace{0.24cm}
 \label{eqC1b}
 \end{eqnarray}
 \end{subequations}
\end{widetext}
We then multiply each term in \eqN{eqC1} by $e^{-ik\Omega_m\tau}\Omega_m/(2\pi)$, where $k\in[-\mathcal{F},\mathcal{F}]$, and integrate them over one modulation period, from $-\pi/\Omega_m$ to $\pi/\Omega_m$. After integration, only one non-zero term remains in each equation. 
Thus, a set of complex-valued linear equations can be obtained:
\begin{widetext}
\begin{subequations}
\label{eqC2}
\begin{eqnarray}
\left[1\!+\!K_c\!-\!(\Omega_f\!+\!k\Omega_m)^{2}\!+\!i2\zeta(\Omega_f+k\Omega_m)\right] x_{1,k}^{F}\!-\!K_c x_{2,k}^{F}\!+\!\frac{K_m}{2} x_{1,k-1}^{F}\!+\!\frac{K_m}{2}x_{1,k+1}^{F}\!=\!\frac{P}{2} \delta_{k,0}\,, \label{eqC2a} \\
\left[1\!+\!K_c\!-\!(\Omega_f\!+\!k\Omega_m)^{2}\!+\!i2\zeta(\Omega_f\!+\!k\Omega_m)\right] x_{2,k}^{F}\!-\!K_c x_{1,k}^{F}\!+\!\frac{K_m}{2}e^{-i\phi}x_{2,k-1}^{F}\!+\!\frac{K_m}{2}e^{i\phi}x_{2,k+1}^{F}\!=\!0\,. 
\label{eqC2b}
 \end{eqnarray}
 \end{subequations}
\end{widetext}
Here, $\delta_{k,0}$, the Kronecker delta, is non-zero (with value 1) if and only if $k=0$; this term determines the location where the external force is applied in the \textit{forward} configuration.
We can write \eqN{eqC2} in matrix form as follows:
\begin{widetext}
\begin{eqnarray}
\label{eqC3}
\begin{bmatrix}
    \vdots & \vdots & \vdots & \vdots & \vdots & \vdots & \vdots & \vdots & \vdots & \vdots\\
    \cdots & A_{-1} & K_m/2 & 0 & \cdots & \cdots & -K_c & 0 & 0 & \cdots\\
    \cdots & K_m/2 & A_{0} & K_m/2 & \cdots & \cdots & 0 & -K_c & 0 & \cdots\\
    \cdots & 0 & K_m/2 & A_{1} & \cdots & \cdots & 0 & 0 & -K_c & \cdots\\
    \vdots & \vdots & \vdots & \vdots & \vdots & \vdots & \vdots & \vdots & \vdots & \vdots\\
    \vdots & \vdots & \vdots & \vdots & \vdots & \vdots & \vdots & \vdots & \vdots & \vdots\\
    \cdots & -K_c & 0 & 0 & \cdots & \cdots & A_{-1} & K_m e^{i\phi}/2 & 0 & \cdots\\
    \cdots & 0 & -K_c & 0 & \cdots & \cdots & K_m e^{-i\phi}/2 & A_{0} & K_m e^{i\phi}/2 & \cdots\\
    \cdots & 0 & 0 & -K_c & \cdots & \cdots & 0 & K_m e^{-i\phi}/2 & A_{1} & \cdots\\
    \vdots & \vdots & \vdots & \vdots & \vdots & \vdots & \vdots & \vdots & \vdots & \vdots\\
\end{bmatrix}
\begin{Bmatrix}
    \vdots\\ y^F_{1,-1}\\ y^F_{1,0}\\ y^F_{1,1}\\ \vdots\\
    \vdots\\ y^F_{2,-1}\\ y^F_{2,0}\\ y^F_{2,1}\\ \vdots\\
\end{Bmatrix}=
\begin{Bmatrix}
    \vdots\\0\\P/2\\0\\ \vdots\\
    \vdots\\0\\0\\0\\ \vdots\\
\end{Bmatrix}
\;, 
 \end{eqnarray}
\end{widetext}
where
\begin{equation}
\label{eqCA_j}
    A_j=1+K_c-(\Omega_f+j\Omega_m)^{2}+i2\zeta(\Omega_f+j\Omega_m)
\end{equation}
for $\;j=0,\pm1,\pm2,\cdots$.
\eqN{eqC3} can be written in a compact notation as:
\begin{equation}
\label{eqC4}
    \underline{\underline{D}}\,\, \underline{y^F}=\underline{p^F}
\end{equation}
Thus, the complex-valued amplitudes of the harmonic terms in the output $x_2^F(\tau)$ can be formally calculated as $\underline{y^F}=\underline{\underline{D}}^{-1}\,\underline{p^F}$.

The size of matrix $\underline{\underline{D}}$ is $4\mathcal{F}+2$ by $4\mathcal{F}+2$.
With the exception of the elements in its main diagonal, first super diagonal, first subdiagonal, $(2\mathcal{F}+1)^{th}$ super diagonal and $(2\mathcal{F}+1)^{th}$ subdiagonal, all the elements in matrix $\underline{\underline{D}}$ are zero.

The only difference between the \textit{forward} and \textit{backward} configurations is the location where the external force is applied. The matrix $\underline{\underline{D}}$ is therefore the same for the two configurations. The force vector $\underline{p^F}$ has only one non-zero element, which is in the $(\mathcal{F}\!+\!1)^{th}$ row. This greatly simplifies the matrix inversion: 
$y^F_{2,p}$, an arbitrary complex-valued amplitude of a harmonic term in $x_2^F(\tau)$, can be calculated from:
\begin{equation}
\label{eqC5}
    y^F_{2,p}=(-1)^{p+1}\,M_{\mathcal{F}+1,3\mathcal{F}+2+p}\,\frac{P}{2\,|\underline{\underline{D}}|}
\end{equation}
where $M_{j_1,j_2}$ is the minor of matrix $\underline{\underline{D}}$ for the element in $j_1^{th}$ row and $j_2^{th}$ column.

Similarly, for the 2-DoF modulated system in the \textit{backward} configuration, the complex-valued amplitudes of all harmonic components in the output $x_1^B(\tau)$ can be formally calculated from $\underline{y^B}=  \underline{\underline{D}}^{-1}\,\underline{p^B}$. 
The force vector $\underline{p^B}$  has only one nonzero element, which lies in the $(3\mathcal{F}+2)^{th}$ row.
$y^B_{1,p}$, an element in $\underline{y^B}$, can then be written as:
\begin{equation}
\label{eqC6}
    y^B_{1,p}=(-1)^{p+1}\,M_{3\mathcal{F}+2,\mathcal{F}+1+p}\,\frac{P}{2\,|\underline{\underline{D}}|}.
\end{equation}

\begin{figure}[hbt]
\includegraphics[scale=0.45]{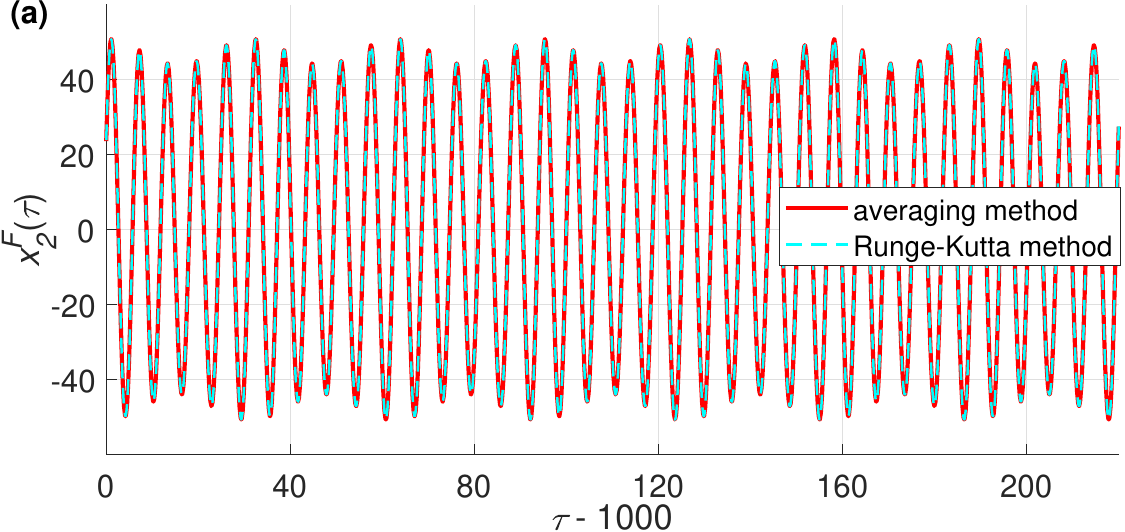}
\includegraphics[scale=0.45]{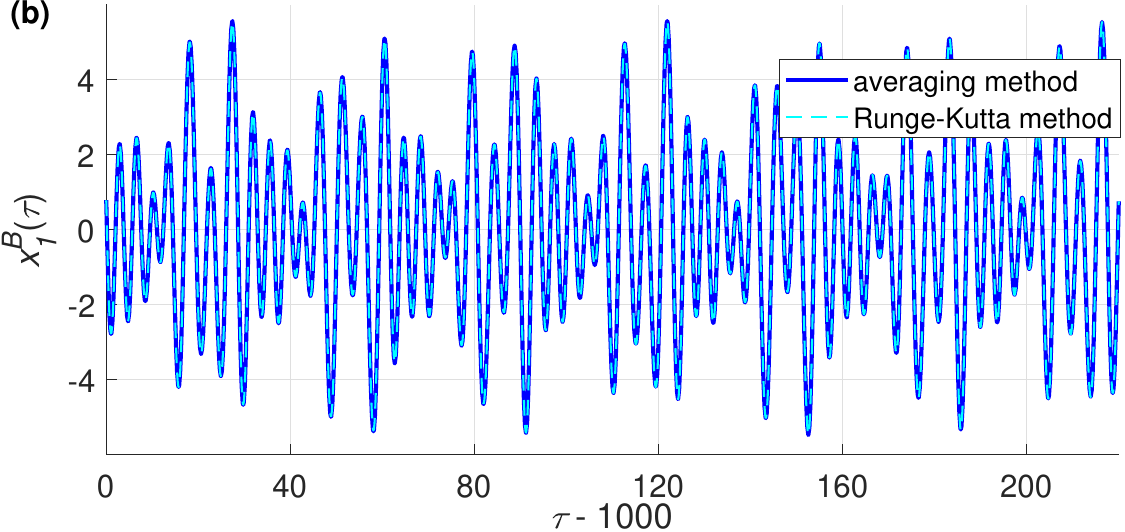}
\caption{\label{fig:Avg-DNI} Comparison of steady-state response calculated using the averaging method (red and blue solid curves) and the Runge-Kutta method (cyan dashed curves). (a) $K_c=0.6$, $K_m=0.1$, $\zeta=0.005$, $\Omega_m=0.2$, $\phi=0.5\pi$, $P=1$, $\Omega_f=1$ and $\mathcal{F}=2$, in \textit{forward} configuration; (b) $K_c=0.7$, $K_m=0.6$, $\zeta=0.005$, $\Omega_m=0.1$, $\phi=0.3\pi$, $P=1$, $\Omega_f=1.33$ and $\mathcal{F}=6$, in \textit{backward} configuration.}
\end{figure}

Fig.~\ref{fig:Avg-DNI} shows the steady-state output displacements calculated for \eqN{eq1} for the following parameters: $K_c=0.6$, $\zeta=0.005$, $\Omega_m=0.2$, $\phi=\pi/2$ and $P=1$.
To validate the predictions made with the averaging method, the outputs of Eq.~\ref{eq:1a} is computed using the Runge-Kutta method. 
The predictions made from the averaging method match very well the results obtained from direct numerical integration.

The same methodology based on the averaging method can be used to obtain the steady-state response of longer discrete modulated systems \cite{NJP2023_Ruzzene,IDETC2023}.

\nocite{*}

\bibliography{arXiv}

\end{document}